\documentstyle[12pt,aaspp4]{article}

\lefthead{Hickson and Mulrooney}
\righthead{UBC-NASA Multi-Narrowband Survey. I.}

\begin{document}

\title{UBC-NASA Multi-Narrowband Survey. I. \\
Description and Photometric Properties of the Survey}
	
\author{Paul Hickson}
\affil{Department of Physics and Astronomy, University of British 
	Columbia, 2219 Main Mall, Vancouver, BC V6T 1Z4, Canada}

\and

\author{Mark~K.~Mulrooney\altaffilmark{1}}
\affil{Department of Space Physics and Astronomy, Rice University, 
	 P.O. Box 1892, Houston, TX 77251, U.S.A.}
	 
\altaffiltext{1}{Project Scientist, Nasa Orbital Debris Observatory} 

\begin{abstract}

Observations, analysis techniques and photometric properties of a
multi-narrowband survey of a 20-deg$^2$ area of sky at $+33\deg$ declination 
are described.  The survey is conducted with the 3-meter
liquid-mirror telescope of the NASA Orbital Debris Observatory using
intermediate-bandwidth filters.  These filters have central wavelengths ranging
from 455 to 948 nm, at intervals of 0.01 in $\log \lambda$, and give a spectral
resolving power of $\Delta\lambda/\lambda \sim 44$.  In two spring
observing seasons we have completed observations in 16 bands.
Our preliminary object catalog contains over $10^5$ detections to a 
typical 50\% completeness limit of $m_{AB} \sim 20.4$.  This paper describes 
the observational techniques, data analysis methods, and photometric properties 
of the survey.  The survey is expected to provide object classifications, 
spectral energy distributions and multi-narrowband redshifts for $\sim 10^4$ 
galaxies and QSOs, for use in studies of galaxy and QSO distributions, 
evolution and large-scale structure. It will also provide photometry and 
spectral classifications for over $10^4$ stars.

\end{abstract}

	
\section{Introduction}

In recent years much progress has been made in understanding the distribution
and evolution of galaxies.  Much of this has been driven by the advent
of redshift surveys.  The CFA surveys (eg.  \cite{dL.86}, \cite{GH89}) provided
the first extensive wide-angle view of the galaxy distribution.  The combined
CFA2 and Southern Sky Redshift Survey (SSRS2:  \cite{dC.94a}, \cite{dC.94b})
covers one third of the sky and contains 14,383 galaxy redshifts with a mean of
$<z> \sim 0.025$.  These surveys reveal a galaxy distribution dominated by
filaments, sheets, and voids, with features comparable in size to the depth of
the surveys.

Fiber-optic surveys, such as the Las Campanas Redshift Survey (LCRS:
\cite{S.96}) and ESO Slice Project (ESP:  \cite{V.97}), have reached greater
depth at the cost of reduced sky coverage and sampling completeness.  The LCRS
extends over 700 deg$^2$ and contains 26,418 galaxy redshifts.  Within the
survey region, approximately 70\% of galaxies having magnitudes in the range
$15.0 < m_r < 17.7$ were measured spectroscopically.  The ESP covers 23.3
deg$^2$ and contains 3342 redshifts to a comparable depth.  These surveys reach
depths of $<z> \sim 0.1$ and reveal structures with sizes of order $50 -
100h^{-1}$ Mpc (the Hubble constant $H = 100h$ km s$^{-1}$ Mpc$^{-1}$).

Multi-slit studies, such as the ESO-Sculptor survey (\cite{B.95}, \cite{BdL95},
\cite{A.97}) and the CNOC surveys (eg.  \cite{YEC96}), sample galaxies to $<z>
\gtrsim 0.3$ over angular areas of a few deg$^2$.  They reach a
magnitude limit $m \sim 21$ with completeness exceeding 50\%, and provide on the
order of $10^3$ redshifts with $<z> \sim 0.3$.  These are complemented by
narrow-field spectroscopic surveys (eg.  \cite{B.88}, \cite{C.90}, \cite{LCG91},
\cite{CSH91}, \cite{L93}, \cite{C.93}, \cite{L.95a}, \cite{K.96}, \cite{M.97})
which cover areas of less than 1 deg$^2$, but reach to faint magnitudes.  For
example, the Canada-France Redshift Survey (CFRS:  \cite{L.95a}, \cite{LF.95},
\cite{L.95b}, \cite{H.95}, \cite{C.95}) contains redshifts for 591 galaxies
having magnitudes in the range $17.5 \leq I_{AB} \leq 22.5$ and has a median
redshift of $<z> \simeq 0.56$.

The deep surveys have provided a wealth of information on galaxy evolution.  The
excess counts of faint blue galaxies detected in early photographic surveys
(\cite{P.79}, \cite{TJ79}, \cite{K80}) were shown not to arise from a new
population of high-redshift objects, but rather from moderate redshift ($z
\gtrsim 0.5$) galaxies undergoing frequent bursts of star formation (for a
review see \cite{KK92}).  The nature of these galaxies, and what becomes of
them, is not yet clear.  Two possibilities are that they disappear from
low-redshift samples by merging (\cite{LS91}, \cite{BEG92}, \cite{CM97}) or by
fading (\cite{BR92}, \cite{BF96}).  If merging dominates, it may be revealed by
studies of the frequency of close pairs of galaxies (\cite{ZK89}).  Recent
studies (eg.  \cite{WFR95}, \cite{P.97}) have reached opposing conclusions;
larger samples are needed to resolve this.  If fading is the dominant effect,
then the progeny of these dwarfs should be detected as an upturn in the local
luminosity function at faint luminosities.  The luminosity function found by the
blue-selected ESP shows evidence for such an upturn (\cite{Z.97}), primarily due
to contributions from blue emission-line galaxies.  The LCRS luminosity function
(\cite{L.96}) does not show the effect, although this may be due to the fact
that the galaxies were selected in the R band.  \cite{D.96} have argued that
neither current wide-angle surveys nor deep surveys sufficiently constrain the
local luminosity function.  Larger samples of faint galaxy redshifts will be key
in resolving these issues.

In order to investigate further the nature of the galaxy population at moderate
redshifts, we have undertaken a survey of $\sim 10^4$ galaxies over a strip of
sky comprising 20 deg$^2$.  The survey is unique both in extent, and in the
nature of the observations.  It provides spectral energy distributions (SEDs)
for all objects, from which are derived morphological classifications and
redshifts.  This paper introduces the survey and discusses its photometric
properties, selection effects and completeness.  The data provide a base for
studies ranging from galactic structure to cosmology.  The main scientific
results will be presented in future papers in the series.  The Object Catalog
will be made available to the community via the World-Wide Web
(http://www.astro.ubc.ca/lmt).

\section{Observations}

The survey observations are conducted using the 3-meter telescope of the NASA
Orbital Debris Observatory (NODO:  \cite{PM97}), located in the Sacramento
mountains of New Mexico, near the town of Cloudcroft.  The site, at an
elevation of 2756 m, is dry and has typically $\sim 120$ photometric nights per
year.  The observatory is located within a 30-minute drive of Sunspot, home of 
the National Solar Observatory (NSO), the Apache Point Observatory, and the 
Sloan Digital Sky Survey Telescope.

The NODO telescope is a zenith-pointing instrument employing a rotating
liquid-mercury mirror.  Recent technological developments at Laval University 
and the University of British Columbia (UBC) have made liquid-mirrors a 
viable low-cost alternative to glass mirrors for large zenith-pointing 
telescopes (eg.  \cite{B.92}, \cite{HGH93}).  The telescope is discussed in
detail by \cite{M98} and \cite{MH98}.  The liquid mirror is similar in design
to that of the UBC-Laval 2.7-meter telescope (\cite{H.94}) and
consists of a rotating parabolic dish covered by a 1.8-mm layer of liquid
mercury.  It is supported by an air bearing coupled to a DC synchronous motor.
The motor is driven by a power supply whose frequency is stabilized to $0.01$
ppm by a crystal oscillator.  The mirror has a focal length of 4.5 meters and
is parabolic to within a fraction of a wavelength.  Off-axis aberrations are
removed by a three-element optical corrector lens located near the prime-focus.
The corrector and detector are supported by a tripod and top-end positioning
system which facilitates mechanical alignment and provides remote control of
focus and detector rotation.

The detector used for survey observations is a $2048 \times 2048$ pixel Loral
CCD.  It is a thick, front-illuminated device, with reasonable response at red
and near-infrared wavelengths but limited blue response.  The pixel size is 15
um square, which gives an image scale of 0.598 arcsec pixel$^{-1}$ in both right
ascension and declination.  The CCD is housed in a thermoelectrically-cooled
dewar.  Heat is removed from the dewar by closed-circuit circulation of a
water-glycol liquid to a cooling unit located on the observatory floor.  The
normal operating temperature of the CCD is -30C, which keeps dark current at a
low level when the CCD is operated in multi-pinned phase (MPP) mode.  During
observations the CCD is scanned continuously in time-delay integrate (TDI) mode
(\cite{MAS80}, \cite{HM84}).  The scan rate is synchronized to the sidereal rate
at the telescope latitude.  The effective integration time is 97.0 s -- the
transit time of images moving across the CCD due to the Earth's rotation.  This
results in a continuous data stream of 90 KB/s, about 0.3 GB/hr.  The CCD is
driven by a Photometrics controller interfaced to a SUN SparcStation-20
computer.  Because the controller is designed for high-speed operation to
support space-debris observations, the read noise is relatively high (28
electrons), and only a 12-bit analog-to-digital converter is provided.  This 
has limited the performance of the detector, particularly at short wavelengths 
where the sky is less bright. We plan to upgrade the system with a
low-noise 16-bit controller before the spring 1998 observing season.

Data acquisition is accomplished by Hickson's {\it Tditool} software running
under the Solaris operating system.  It provides interactive control of
data-acquisition parameters and a continuous display of image data.
Each CCD line is written directly to an 8-mm data tape and to the
computer display.  A 32-MB buffer memory prevents
any data loss due to interrupt-response delays by the operating system.  The
display is provided with zoom and contrast controls which facilitate
monitoring of focus and image-quality during the observations.  One 8-mm tape
has sufficient capacity for a single night of data.

The telescope is equipped with a holder for a single 4-inch-diameter glass
filter, located between the corrector lens and the detector window.  The filter
is inserted at the beginning of the night, and is used for the entire
night's observations.  The filters are selected from UBC's set of 40
intermediate-bandwidth interference filters designed by Hickson specifically 
for multi-narrowband imaging.  They have central wavelengths at a 
uniform logarithmic spacing of 0.01 and constant bandwidth of 0.02 in 
$\log \lambda$.  Because of the declining sensitivity of the CCD at long
and short wavelengths, only 33 filters are used in the survey.  These have
central wavelengths in the range $455 - 948$ nm, with corresponding bandwidths
ranging from 17.6 to 39.1 nm.  The logarithmic sampling interval of the filter
set provides equal redshift resolution at all wavelengths.  The individual
filter bandwidths naturally increase in proportion to the central wavelength,
and are set at twice the sampling interval in order to prevent aliasing effects
when observing emission-line objects.  The characteristics of the filters are
described in Table 1.  Their transmission curves are shown in Figure 1.

Observations were conducted during the months of April-June in 1996 and 1997.  
In 60 nights of observing we obtained data in 16 wavelength bands spaning
the entire range of sensitivity of our detector.  Almost all nights were 
photometric.  We obtained a minimum of three nights of observations 
with each filter.  This was done in order to allow the rejection of spurious 
events such as cosmic-ray hits, to provide an independent estimate of the 
photometric accuracy, and to improve the signal-to-noise ratio of the 
detections. In some bands additional nights were obtained because of poor 
seeing or clouds. Table 2 provides a summary of the bands observed, the
number of nights, and the ranges of right ascension for which we have data.  

\section{Data analysis}

The large quantity of data generated by the survey observations requires a
dedicated data analysis facility and considerable software development.  The
data tapes obtained in the survey were analyzed at UBC using a software package
{\it Lmtphot}, written explicitly for the analysis of liquid-mirror telescope 
observations.  The package consists of interrelated programs which provide
an automated path from the raw data tapes to the final catalog of object 
photometry.  This section gives a brief description of these steps; more 
details can be found in
\cite{H97}.

\subsection{Preprocessing}

Preprocessing steps consist of dark/bias subtraction, flat-field correction, and
sky subtraction.  With TDI operation, each image pixel results from integration
over the entire length of a CCD column.  Because of this, variations in CCD
pixel response are greatly diminished.  The corrections for both dark current
and flat-field are one-dimensional.  The dark correction is determined by
averaging many lines of TDI data taken with the CCD covered.  Flat-field
exposures are obtained on cloudy moonlit nights.  By median filtering several
hours of observations, any star images and cosmic rays are removed.  After dark
and flat-field correction, consecutive CCD lines are grouped into overlapping
2048-line blocks.  Sky subtraction is then performed by subtracting the mode of
each row and column.  Because of the high degree of uniformity in TDI images,
this technique works very well.  Systematic variations in the background 
after sky subtraction are typically less than 0.02\%.

\subsection{Object detection and photometry}

Object detection and photometry is performed on the individual data blocks
described above.  In order to reduce noise, object detection is done on a
smoothed copy of the image.  The smoothing is normally accomplished by means of
a $5 \times 5$ pixel ($3 \times 3$ arcsec) boxcar filter, which provides a
5-fold reduction in background noise.  Other filter sizes may be selected for
special programs.  In each block, the mean and standard deviation of the
background is determined using iterative outlyer rejection to eliminate star
images.  All pixels having values more than 2.5 standard deviations above the
mean are noted and all connected sets of such pixels are found.  The boundary
of each such set is taken to be the detection isophote of an object.  For each
object the first three moments of the intensity distribution within the
detection isophote are determined.  The zero-order moment gives the isophotal
flux; the first-order moments give the centroid of the image, from which
coordinates are computed.  The second-order moments determine the
moment-of-inertia tensor.  From its eigenvalues the major and minor axis
diameters and position angle of the equivalent Gaussian ellipsoidal intensity
distribution are found.  All photometric measurements are done on the
original unsmoothed image, within the boundary determined from the smoothed
image.  The smoothed image is used only for object detection and the
delineation of the detection isophote.

It is well-known that raw isophotal magnitudes become seriously biased for 
faint objects because an increasing fraction of the light from the object falls
outside the detection isophote.  On the other hand, adding flux from outside
the isophote increases the noise of the measurement because the signal-to-noise
ratio of exterior pixels is low.  We choose instead to apply a correction to
the isophotal magnitudes based on the flux and mean intensity within the
detection isophote. This correction is computed by assuming that the 
relationship between the intensity of any isophote and the flux within that 
isophote is the same as for a Gaussian intensity profile. Although disk
galaxies might be better represented by an exponential profile, a Gaussian
provides a good approximation to the seeing-degraded profiles of the faintest 
galaxies in our images, for which the magnitude correction is significant.
The method was tested by applying it to realistic simulations of artificial 
images having scale, resolution and noise comparable to the actual data. The 
corrected-isophotal technique was found to provide an unbiased estimate of 
the true total magnitude and to have lower noise than alternative esimators. 
Details of the correction procedure and the simulations are described by 
\cite{H97}.
 
The advisability of using total magnitudes to
estimate spectral energy distributions is debatable because the flux in
different bands is not determined from exactly the same areas.  (See
\cite{KK92} for a discussion of the difficulties of measuring colors of faint
galaxies).  On the other hand the use of a common fixed aperture has its own
problems.  Unless the aperture is much larger than the seeing disk, seeing
variations will cause significant changes in the fraction of flux within the
aperture.  However, such large apertures generally result in unacceptably-high
levels of sky noise.  In addition, noise fluctuations affect the alignment of
the apertures, so the areas of measurement are in fact never exactly the same.

In order to identify and separate objects whose images overlap, the object
detection algorithm is run repeatedly using increasingly-bright thresholds.
At each stage the object list is examined for multiple detections within
the isophotal boundary of every object found in the previous iteration. In
such cases the isophotal and total fluxes of the ``parent'' object, determined 
from the original ($2.5\sigma$) isophote, is divided amongst the ``children'' 
in proportion to their isophotal fluxes at the time that their images separate.
This recursive technique is similar to that used by other photometric packages
and is very successful at separating blended images. For our data, the
fraction of objects affected by blending is typically of order 1\%.

The photometry program produces a list of all objects detected in a night's
observations, with instrumental magnitudes and positions, estimated errors,
image parameters and the seeing FWHM, determined from star images.  Initial
(instrumental) coordinates are determined by applying corrections for
aberration, nutation to the Cartesian coordinates of the image centroids, then
precessing these values to a standard epoch.  The preprocessing, object
detection and photometry is the most time-consuming part of the analysis,
requiring about 4 hours of CPU time on a Sun Ultra-I computer to process a
3-GB data tape.  The process produces a photometry file, with typical size
$\sim 10$ MB, for each night of observations.

\subsection{Astrometric and photometric calibration}

An astrometric and approximate photometric calibration is obtained from data in
the Hubble Space Telescope Guide-Star Catalog (GSC: \cite{GSC1}, \cite{GSC2},
\cite{GSC3}).  The calibration process consists of matching stars in the GSC
with those in the photometry file, then applying appropriate corrections to the
instrumental positions and magnitudes.  The matching is done iteratively,
beginning with the brightest stars to obtain a rough fit, then improving the fit
by using the more numerous fainter stars, one magnitude interval at a time.  At
each level, five free parameters are fit.  These are the coefficients of a
linear regression in right ascension and declination and a fixed magnitude
correction.  The fitting error is minimized in 5-dimensional parameter space by
means of the simplex algorithm.  The technique is robust and converges to the
correct solution in a relatively short time.  It typically takes about 5 min to
calibrate an entire photometry file.  The resulting coordinates have typical RMS
errors of $\lesssim 1.0$ arcsec in both right ascension and declination.  We
believe that the technique is capable of greater accuracy, but we are presently
limited by distortion in the telescope's optics.  The magnitude fitting is 
necessary to correctly match the target and GSC stars.  A constant is added 
to the instrumental magnitudes of the target stars so that their mean 
magnitude matches that of the listed V-band magnitudes of the corresponding 
GSC stars.

After astrometric calibration, a more-accurate photometric calibration is made
using spectrophotometric standard stars in the survey field.  As only two stars
in our survey area have published spectrophotometry (HZ 21 and BD+332642) a
program to establish secondary spectrophotometric standards in the LMT survey
area was undertaken at Kitt Peak national Observatory (KPNO).  Using the
2.1-meter telescope and Goldcam spectrograph, 22 stars were observed in the
region of sky observed by the NODO telescope.  The resulting spectrophotometry
is reported in a separate paper (\cite{HM98}).  These stars pass through the
NODO telescope field at approximately 30 min intervals, providing an accurate
calibration for all our wavelength bands.  For each band, the product of the
filter transmission curve and standard star specific flux is integrated to
provide the magnitude zero point for the band.  In order to correct for any
slow variations in transparency during the night, a second-order polynomial is
fit to the zero points obtained from the standard stars and applied to the
instrumental magnitudes to give the calibrated magnitudes for all objects.

We use the AB magnitude scale (\cite{O74}) defined by
\begin{equation}
  m_\nu = 56.10 - 2.5\log f_\nu
\end{equation}
where $f_\nu$ is the specific flux, in W m$^{-2}$ Hz$^{-1}$, averaged over the
filter bandpass.  As our filters have relatively narrow bandwidths, this is a
good approximation to the monochromatic flux at the central wavelengths of the
filters (Table 1).

\subsection{Merging and cataloging of object data}

The next step is the merging of calibrated photometry files from all nights in
which observations were made with the same filter.  The program identifies all
objects which are detected with position errors of less than 3.5 arcsec and
magnitude differences of 1.0 or less.  In order to reject cosmic rays and
spurious detections of noise, we require that an object be detected on more 
than one night. 3.5 arcsec corresponds to $2.5\sqrt{2}$ times the
typical astrometry error. The chance of an object having a position
error larger than this is about 0.01. At the lowest galactic latitudes
reached by the survey, the  density objects, most of which are stars,
is of order $10^{-3}$ arcsec$^{-2}$. The probability of finding a second 
object within 3.5 arcsec of a given object is therefore of order 
$1-\exp(-0.006\pi) \simeq 0.02$. Thus the 3.5-arcsec criterion leads
to roughly comparable rates of dropouts and contamination in the
most crowded area of the survey. 

For each night, the random error in the magnitude is estimated from
the object counts, isophote area and background variance.  The mean magnitude
is then determined, weighting the magnitudes for each night by the reciprocal
variances.  The same weights are used when computing the mean values of the
other photometric and astrometric parameters for each object.  The final
photometric errors are computed as follows:  For each object, the variance of
the mean is computed from the individual magnitude variances.  This provides an
estimate of the random noise.  However, systematic errors can occur due to
imperfect correction of extinction variations.  The total error (random plus
systematic) is estimated by computing directly the variance of the magnitudes
obtained on different nights.  The larger of these two error estimates is
adopted.

This procedure produces a single photometry file for each wavelength band.
These files are then merged to form a single file of SEDs for each object.
Objects whose positions agree to within 3.5 arcsec are assumed to be the same,
and the individual magnitudes, in the different bands, are entered in the
corresponding fields of the object data in the output file.  The result is a
set of magnitudes, in as many as 33 bands, for each object.

\section{Performance of the telescope}

The image quality of the liquid-mirror telescope is primarily limited by the
seeing at the site.  The median FWHM of our star images is 1.4 arcsec, although
images as small as 0.9 arcsec were occasionally recorded.  While the present
corrector does provide good image quality over the entire 0.33-deg field, it 
does not remove distortion.  This is a problem for TDI observations because 
the images do not track in straight lines at a constant rate.  The field 
distortion is in fact more serious than the usual star-trail curvature effects 
which are present due to the non-zero observatory latitude (\cite{GH92}, 
\cite{ZSB96}).  As a result, the images suffer from a varying degree of image 
smear which, while small at the field center becomes $\sim 2$ arcsec near the 
north and south edges.  Because of this, the survey area was restricted to the 
central 75\% (16 arcmin) of the CCD field of view.  We plan to upgrade the 
corrector with one designed for TDI observations before the spring 1998 
observing season.

The combined throughput of the atmosphere, telescope, corrector and CCD was
determined from the observed flux of the standard stars, and is shown in Figure
2 as a function of wavelength.  The throughput is $\sim 18$\% between 650 and
800 nm and declines rapidly at wavelengths shorter than 500 nm and longer than
900 mn due to the falling CCD response.  The overall efficiency and shape is
consistent with expectations.  The useful wavelength range with this CCD is
approximately $450-950$ nm.

One area of concern with liquid mirrors is the effect of wind.
Under calm observing conditions, fluctuations in the rotation
period of the mirror are ~ 10ppm, due primarily to thermal gradients
in the dome area. Larger fluctuations are observed on windy nights,
due to wind gusts affecting the mirror. Although the telescope is well
shielded inside the observatory's 50' diameter dome, useful
observations are not possible when exterior winds exceed ~12 m s$^{-1}$.
With such high wind speed, variations in mirror rotation speed of order 
30 ppm are seen.  The image quality is degraded (star images have 
FWHM $\sim 2.4$ arcsec), but at least part of this may result from poor 
atmospheric seeing associated with high wind speed.

\section{Properties of the survey}

The coordinates and area of the survey region are indicated in Table 3.  Since
the NODO telescope is a zenith-pointing instrument, the region of sky surveyed
is determined by the observatory latitude and the time of observations.  Our
spring observing season, characterized by good weather at the site, typically
extends from early April until late June.  Although some observations were made
with right ascension as early as 9 hrs, the region of overlapping coverage
extends approximately from 12-19 hrs.  In order to avoid the high stellar
density near the galactic plane, which would make accurate photometry
difficult, we chose to end the survey region at 18:00 hrs right ascension.  The
total area is 20.13 sq deg.

The survey region extends over a wide range of galactic latitude, $10\deg > b >
85\deg$, and contains over $10^5$ stars and galaxies. Because of image
distortion introduced by the telescope corrector, we do not attempt to
distinguish stars from galaxies on the basis of image structure, although
it should be possible to do this using images obtained with the new corrector.
However, unlike conventional photometric surveys, stars and galaxies can be
distinguished by their different spectral signatures. In their study
of the multi-narrowband imaging technique, Hickson, Gibson and Callaghan 
(1994) simulated both stars and galaxies and included 81 stellar
templates, as well as galaxy templates, in the analysis program. For
signal-to-noise ratios of 5 or more, galaxies were rarely confused with 
stars. Similar results have been found by Peri, Iovino and Hickson (1997) 
in simulations of QSO spectra. In the spectral analysis of the catalog,
objects will be compared with both galaxy and stellar templates, in order
to separate stars and galaxies, provide spectral classifications for
both and estimate galaxy redshifts. The resulting stellar data set will 
be useful for studies of stellar populations and galactic structure.

As an illustration of the photometric accuracy of the survey, we plot in Figure
3 the magnitude differentials from two different nights, at 752 nm, as a
function of magnitude. The curve shows the RMS magnitude difference
which rises from $\sim 0.04$ for $m < 16$ to 0.52 at $m = 21$. From this we
can estimate the standard error in our magnitudes, which are based
on the mean of typically three measurements (from three nights of 
observations). The mean of three magnitude measurements is smaller than the 
RMS difference between any two individual measurements by a factor of 
$\sqrt{6} \simeq 2.45$. Thus, we expect our mean
magnitudes to have typical errors ranging from 0.02 to 0.2 mag. The
absense of magnitude differentials greater than 1.0 is a result of
the 1-magnitude limit imposed by the object pairing algorithm. From the
figure it is evident that the number of objects missed because of this 
becomes significant only at magnitudes comparable to the 50\% completeness 
limit.

\section{Selection Effects and Completeness}

Before the survey data can be used for statistical analysis, a complete
characterization of the selection effects is needed.  Our photometry algorithm
finds all light distributions which exceed a surface-brightness threshold over a
minimum number of contiguous pixels.  The detected objects will therefore be
limited by surface brightness as well as by apparent magnitude.  In addition,
there will be an angular separation limit below which objects are not
individually distinguished.

In order for an object to be detected, the smoothed surface brightness in its
image must exceed the detection threshold for a minimum of 5 contiguous pixels.
This corresponds to a minimum area $A_m = 1.788$ arcsec$^2$.  Thus, if $f$ is
the flux within the detection isophote and $\bar{i}$ is the mean intensity
within this isophote, the first selection criterion is
\begin{equation}
  f/\bar{i} \geq A_m 
\end{equation}

Clearly, for the object to be detected at all, we require 
\begin{equation}
  \bar{i} > i_m~, 
\end{equation}
where $i_m$ is the detection threshold intensity.  Because object detection is
done on a smoothed version of the image, these selection effects
refer to the average surface brightness of the object over the smoothing area
(normally $3 \times 3$ arcsec).

A third selection effect occurs because we require a minimum signal-to-noise
ratio $\zeta$.  The noise has contributions from the Poisson noise in the image
and the background, hence,
\begin{equation}
  f > \zeta^2(g + \sigma^2/\bar{i})~,
\end{equation}
where $\sigma^2$ is the background noise variance and $g$ is the system
gain (the signal produced by a single photoelectron). Equations (2 - 4)
represent the three photometric selection criteria of the survey.

In order to maximize both the number of objects detected, and the accuracy with
which they are measured, the surface brightness of the detection isophote is
set at as low a level as the background noise will permit.  Thus, $i_m =
\epsilon n^{1/2}\sigma$ where $n$ is the number of pixels in the smoothing
kernel and, typically, $\epsilon = 2.5$.  The background noise is generally
dominated by light from the sky, which varies substantially with lunar phase.
Thus, the selection criteria (2) and (3) are local rather than global.  The
values of $\sigma$, for each object, are recorded in the individual photometry
files, along with $n$, $\epsilon$, $\zeta$, and $g$.  From these the selection
criteria can be determined for each object.

Although the selection criteria are local, global averages can be made
for any subset of data. This is illustrated in Figure 4, in which the
mean surface brightness
\begin{equation}
  \mu = m - 2.5\log A~,
\end{equation}
where $A$ is the area of the detection isophote, is plotted vs $m$ for objects
in the range $12.0 - 12.1$ hrs in the 752 nm band.  The three solid lines
indicate the respective selection criteria, Equations (2 - 4).  There is good
agreement between the shape of the region delineated by the lines and the
boundaries of the data points.

The magnitude limit for any portion of the survey data can be estimated
from the surface brightness of the detection isophote, the minimum
area required for detection, and the seeing FWHM. This is a function
of sky brightness and atmospheric conditions, and so can change on short 
timescales. Since all necessary data are recorded in the catalog, it
is possible to compute the limiting magnitude for any time and 
wavelength band.

The completeness limit can be esimated by examining the counts $N(m)$ of objects
whose magnitude is less than or equal to $m$.  At high galactic latitude, the
counts at the faintest magnitudes are dominated by galaxies, which have a
linear $\log N$ vs $m$ relation.  We calculate the 50\% completeness limit
$m_{1/2}$ as the magnitude at which the observed counts fall below a linear
extrapolation of the $N(m)$ relation by a factor of two.  Figure 5 shows the
$N(m)$ plot for the 752 nm band.  The 50\% completeness limits for the
various wavelength bands range from 19.0 to 21.1; the median value is 20.4.

\section{Discussion}

This paper introduces the UBC-NASA Multi-Narrowband Survey, which began in
1996 and is now $\sim 50$\% complete.  Observations have been made in 16 
wavelength bands and are continuing in order to enlarge the data set to 33 
bands. Our preliminary catlog is reasonably complete to a magnitude of 
$m_{AB} \sim 20.4$ a surface-brightness of $\mu_{AB} \sim 23.5$. In addition
to the narrowband data, we also obtained 15 nights of observations
using broad-band B, V, R and I filters, which are currently being reduced.
These data will compliment the narrowband data and allow us to reach 
fainter magnitude limits with lower spectral resolution.

These data provide a base for studies of the galaxy space density and luminosity
function to $z \sim 0.5$, galaxy spectral evolution and large-scale structure.
Simulations (\cite{HGC94}, \cite{CB95}, \cite{CH97}) indicate that galaxy
redshifts can be obtained with an accuracy of $\Delta z \sim 0.03$ with the full
set of 33 bands.  This should suffice to detect the presence of structures on
scales reported by \cite{Br.90} if they exist.  Also, simulations indicate that,
by means of topological measures, such data can also serve to distinguish
competing models of galaxy formation by means of topological measures
(Brandenberger, priv.  comm.).

With the data already in hand, redshifts with accuracy $\Delta z \sim 0.05$
are achievable. The object catalog will be refined as additional observations
are obtained. The image database will also allow targeted programs
not possible from the catalog alone. For example, by increasing the 
degree of smoothing, the surface-brightness limit can be extended to
$\mu \sim 26$ in order to detect and study low-surface-brightness galaxies.
The very high degree of background uniformity provided by the TDI
imaging technique make this data set particularly valuable for the
study of such objects, and a study is presently underway (Mulrooney 1998).

The catalog is expected to contain $\sim 10^3$ QSOs (\cite{Z.92}).  Simulations
(\cite{PIH97}) indicate that QSO's can be reliably detected, and distinguished
from stars, at signal-to-noise ratios $\gtrsim 7$ from the survey data.  While
our survey area is less than a factor of two larger than that of the Durham-AAT
survey (\cite{B.90}), which reaches a comparable magnitude limit, our
multi-narrowband technique is relatively free from selection biases and should
be sensitive to quasars over a wide range of redshifts.

The several hundred thousand stars contained in the catalog will be of value
for studies of stellar populations and galactic structure. The multi-narrowband
photometry will provide accurate spectral classifications as well as magnitudes
and colors for most of these stars. In addition, the geometry of the survey 
area probes a relatively large range of galactic latitude.

This survey will compliment, and provide experience for a similar
survey to be conducted at $+49$ deg declination using a 6-m 
liquid-mirror telescope (the LZT project: \cite{H.97}). This instrument,
the successor to the UBC-Laval 2.7m, is currently under construction and
is expected to see first light in 1998. It will use the UBC filter
set and should reach $m_{AB} \sim 23$ and measure over $10^5$ galaxies.

\acknowledgments

We are grateful to NASA for making the NODO telescope available to us for
astronomy, on a non-interference basis with debris observations, to NSO for
invaluable technical and administrative support, and to KPNO for allocation of
observing time and technical support on the 2.1-meter telescope.  PH thanks
NASA and NSO for hospitality during visits to JSC, Sunspot and the NODO.  We
thank B. Gibson for digitizing the transmission curves of the filters, F.
Peri for initial work on astrometry calibrations, and V. de Lapparent,
R. Cabanac and E. Borra for discussions.  PH is supported by grants from 
Natural Sciences and Engineering Research Council of Canada.  The Orbital 
Debris Observatory is funded by NASA.

\clearpage

\begin{deluxetable}{lrrrrrr}
\footnotesize
\tablecaption{Filter Specifications. \label{tbl1}}
\tablewidth{0pt}
\tablehead{
\colhead{ID} & 
\colhead{$\lambda_0$\tablenotemark{a}} & 
\colhead{$\Delta\lambda$\tablenotemark{b}} & 
\colhead{$\log(\nu_0)$\tablenotemark{c}} & 
\colhead{$\Delta\log(\nu)$\tablenotemark{d}} & 
\colhead{$t_0$\tablenotemark{e}} & 
\colhead{$W$\tablenotemark{f}}
} 
\startdata
948 & 947.7 & 39.08 & 14.5003 & 0.019 & 0.933 & 36.43 \nl
925 & 924.5 & 40.04 & 14.5111 & 0.019 & 0.928 & 36.96 \nl
906 & 906.3 & 35.32 & 14.5198 & 0.018 & 0.900 & 31.71 \nl
883 & 883.1 & 41.28 & 14.5311 & 0.021 & 0.924 & 38.10 \nl
868 & 867.9 & 35.10 & 14.5388 & 0.018 & 0.952 & 33.38 \nl
844 & 843.8 & 35.58 & 14.5509 & 0.019 & 0.932 & 33.09 \nl
825 & 824.8 & 33.67 & 14.5608 & 0.018 & 0.950 & 31.96 \nl
806 & 805.9 & 34.62 & 14.5709 & 0.019 & 0.936 & 32.27 \nl
788 & 787.5 & 33.31 & 14.5809 & 0.019 & 0.927 & 30.85 \nl
770 & 769.6 & 31.86 & 14.5910 & 0.018 & 0.937 & 29.79 \nl
752 & 752.4 & 33.25 & 14.6008 & 0.019 & 0.955 & 31.72 \nl
735 & 734.7 & 32.17 & 14.6111 & 0.019 & 0.940 & 30.17 \nl
719 & 718.7 & 30.54 & 14.6208 & 0.019 & 0.954 & 29.13 \nl
704 & 704.4 & 29.88 & 14.6293 & 0.019 & 0.930 & 27.78 \nl
688 & 688.0 & 29.20 & 14.6397 & 0.019 & 0.936 & 27.30 \nl
671 & 671.3 & 29.08 & 14.6503 & 0.019 & 0.933 & 27.10 \nl
655 & 654.6 & 27.99 & 14.6612 & 0.019 & 0.930 & 26.03 \nl
641 & 641.1 & 23.98 & 14.6705 & 0.016 & 0.919 & 21.99 \nl
629 & 628.7 & 26.39 & 14.6789 & 0.018 & 0.952 & 25.10 \nl
614 & 613.7 & 23.62 & 14.6893 & 0.018 & 0.910 & 21.45 \nl
598 & 597.6 & 24.31 & 14.7010 & 0.018 & 0.717 & 17.41 \nl
586 & 585.6 & 23.10 & 14.7099 & 0.018 & 0.720 & 16.63 \nl
571 & 571.1 & 21.71 & 14.7207 & 0.017 & 0.750 & 16.26 \nl
557 & 557.0 & 21.35 & 14.7314 & 0.017 & 0.707 & 15.05 \nl
545 & 545.1 & 21.00 & 14.7409 & 0.017 & 0.726 & 15.22 \nl
533 & 532.7 & 22.76 & 14.7505 & 0.019 & 0.730 & 16.59 \nl
519 & 519.0 & 22.72 & 14.7609 & 0.022 & 0.679 & 15.38 \nl
510 & 510.2 & 22.36 & 14.7698 & 0.019 & 0.689 & 15.39 \nl
498 & 498.1 & 21.91 & 14.7798 & 0.019 & 0.670 & 14.66 \nl
486 & 486.0 & 20.22 & 14.7904 & 0.019 & 0.752 & 15.18 \nl
476 & 475.6 & 19.30 & 14.7998 & 0.018 & 0.690 & 13.31 \nl
466 & 465.9 & 18.48 & 14.8090 & 0.018 & 0.673 & 12.42 \nl
455 & 454.5 & 17.67 & 14.8196 & 0.018 & 0.632 & 11.17 \nl
\enddata
\tablenotetext{a}{mean wavelength (nm), from transmission curve}
\tablenotetext{b}{bandwidth (nm): equivalent width/central transmission}
\tablenotetext{c}{log of central frequency (Hz): c/mean wavelength}
\tablenotetext{d}{log frequency bandwidth: $0.434 \times $bandwidth/mean wavelength}
\tablenotetext{e}{central transmission}
\tablenotetext{f}{equivalent width (nm): integral of transmission curve}
\end{deluxetable}

\clearpage

\begin{deluxetable}{lrr}
\tablecaption{Observations \label{tbl2}}
\tablewidth{0pt}
\tablehead{
\colhead{Filter} & 
\colhead{Nights} & 
\colhead{ra range}
}
\startdata
455 & 3 & 11:05 - 19:35 \nl
510 & 3 & 11:32 - 19:30 \nl
557 & 6 & 09:42 - 19:00 \nl
571 & 5 & 08:24 - 19:01 \nl
598 & 2 & 11:43 - 19:30 \nl
629 & 3 & 10:08 - 19:01 \nl
655 & 3 & 10:10 - 19:31 \nl
688 & 3 & 09:23 - 19:03 \nl
704 & 3	& 11:49 - 19:05 \nl
752 & 3 & 10:30 - 19:15 \nl
788 & 4 & 09:57 - 18:58 \nl
806 & 3 & 12:21 - 19:40 \nl
844 & 5 & 10:20 - 19:08 \nl
868 & 7 & 10:40 - 19:35 \nl
906 & 4 & 12:30 - 19:05 \nl
948 & 3 & 10:23 - 19:03 \nl
\enddata
\end{deluxetable}

\clearpage

\begin{deluxetable}{lr}
\tablewidth{0pt}
\tablecaption{Survey Parameters \label{tbl3}}
\tablehead{
} 
\startdata
ra range (2000):       & 12 00 00 - 18 00 00 \nl
dec range (2000):      & 32 52 00 - 33 08 00\nl
central dec (2000):    & 33 00 00 \nl
dec width (arcmin):    & 16.00 \nl
ra length (arcdeg):    & 75.48 \nl
area (sq deg):         & 20.13 \nl
pixel size (arcsec):   & 0.598 \nl
wavelength range (nm): & 455 - 948 \nl
spectral resolving power: & 44 \nl
magnitude limit (typical): & 20.4 \nl
\enddata
\end{deluxetable}

\clearpage

\figcaption[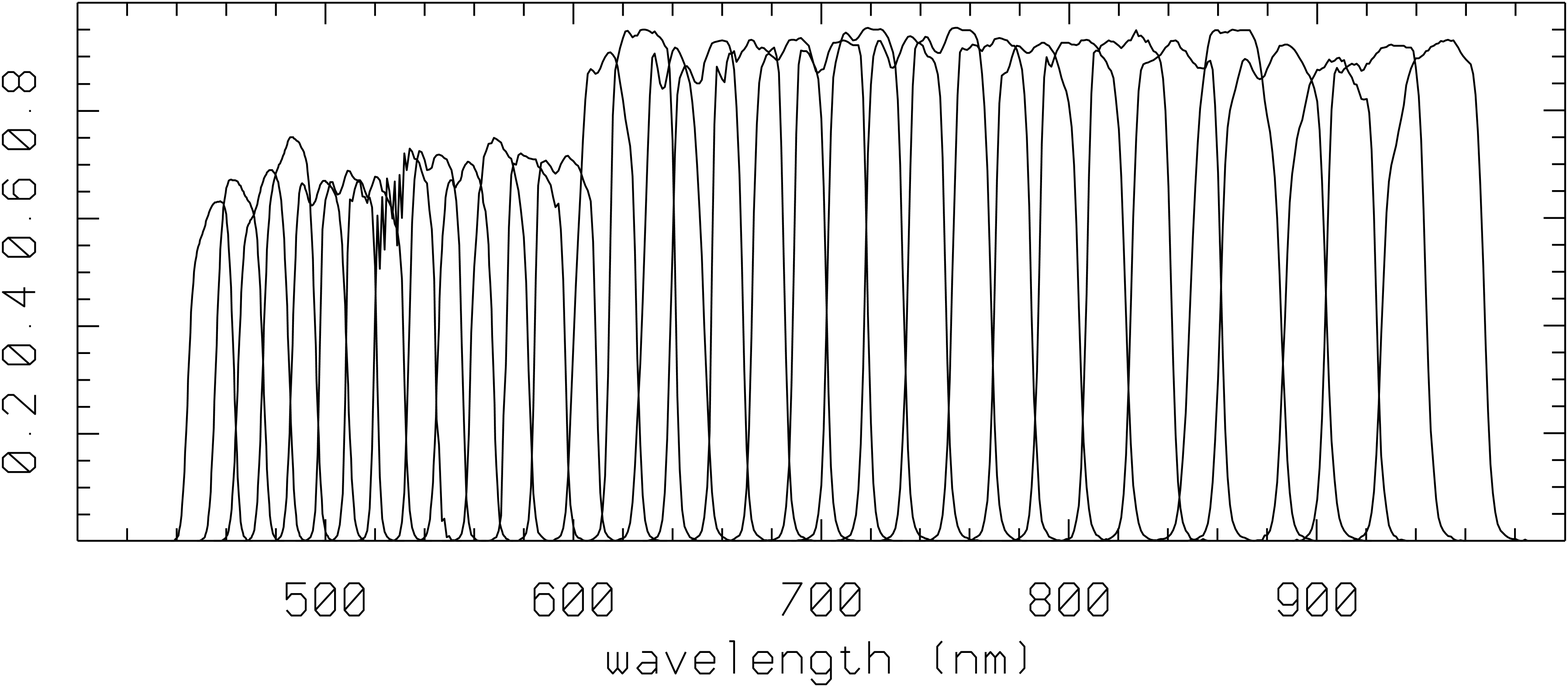]{Transmission curves of the filters.}

\figcaption[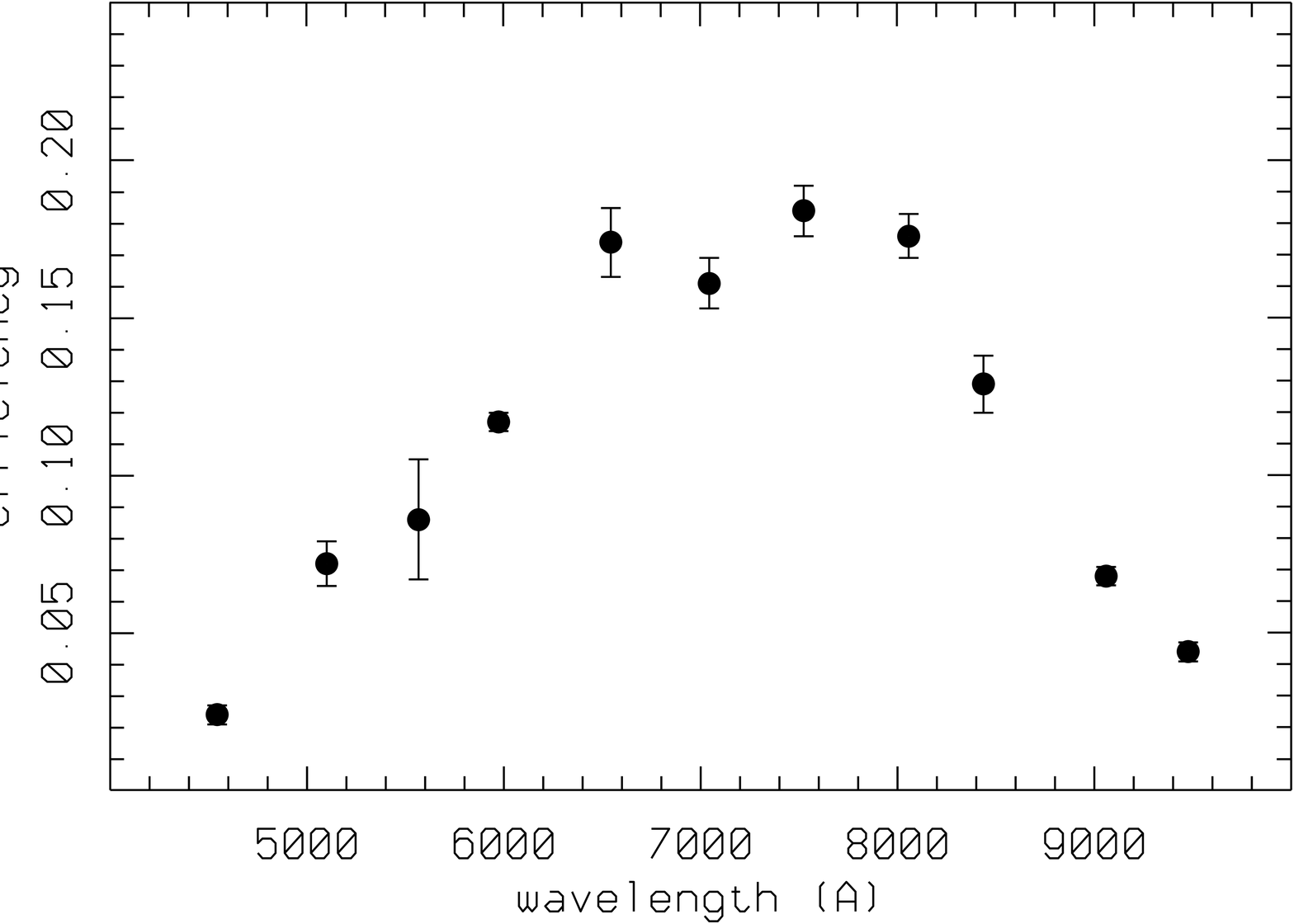]{System throughput. The product of atmospheric transmission,
primary-mirror reflectivity, corrector transmission and CCD quantum 
efficiency is plotted vs. wavelength.}

\figcaption[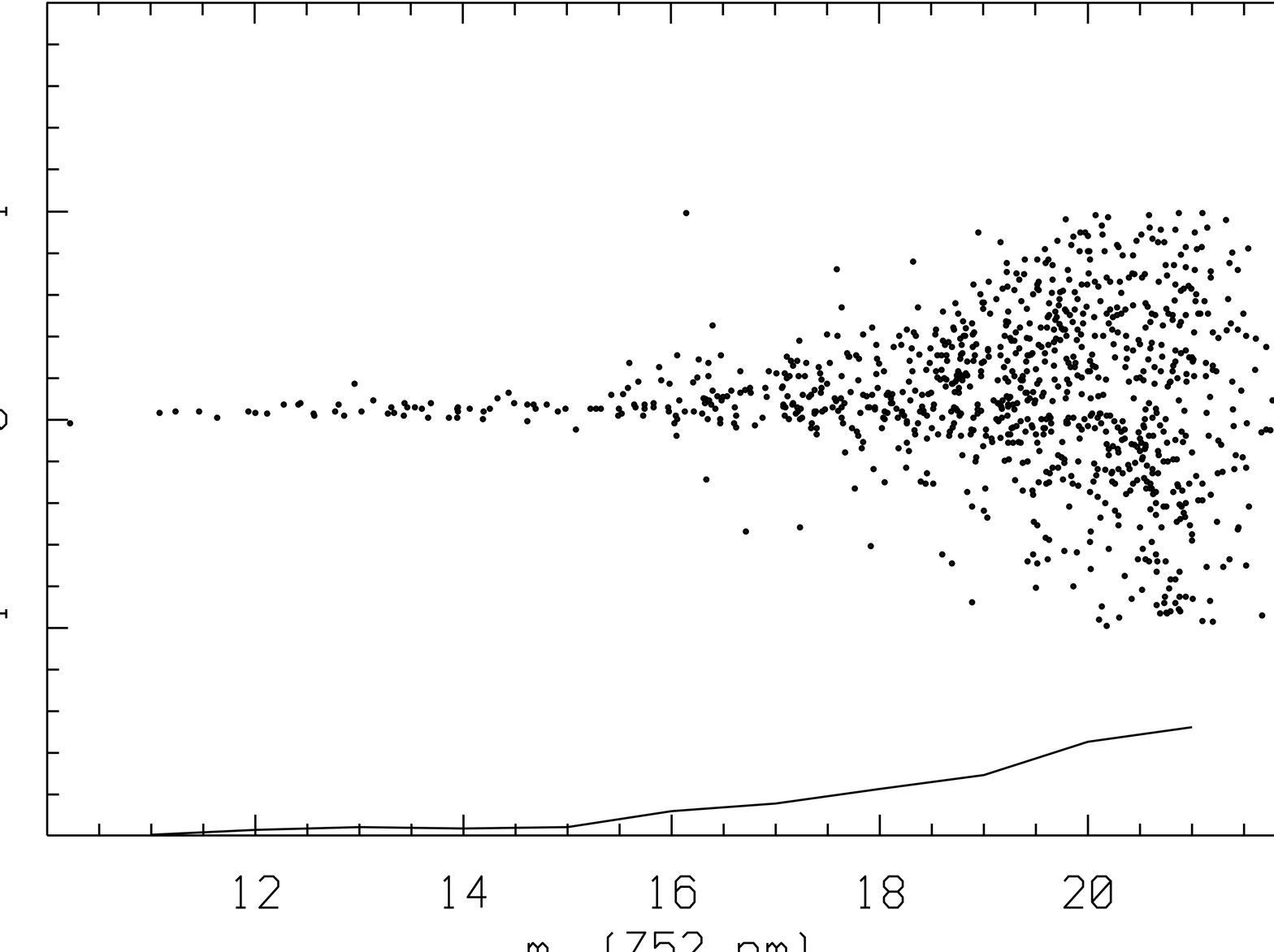]{Photometric residuals. The difference between magnitudes 
measured on two successive nights, in the 752 nm band, is plotted vs. the mean
magnitude. The curve shows the RMS magnitude difference, which rises
from $\sim 0.04$ for $m < 16$ to 0.52 at $m = 21$. The standard
error in our final magnitudes (which result from the mean of typically three 
measurements) is smaller than this RMS difference by a factor of 
$6^{1/2} \simeq 2.45$.}

\figcaption[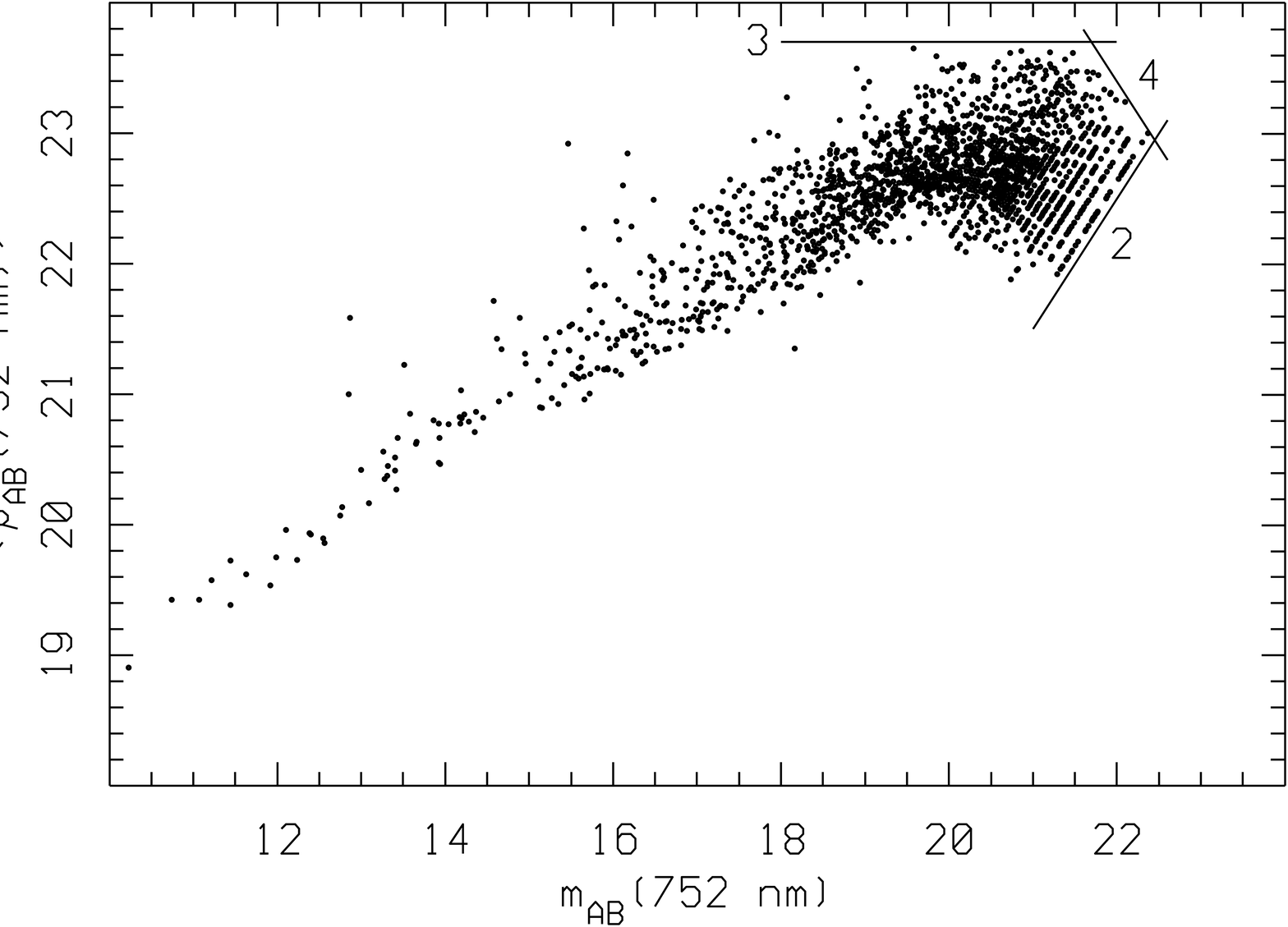]{Selection effects. Mean surface brightness within the 
detection isophote is plotted vs. total magnitude, in the 752 nm band, for 
objects with right ascension in the range $12.0 - 12.1$ hrs.  The solid 
lines, labeled with numbers corresponding to equations in the text,
indicate the three selection effects of surface area (2), surface 
brightness (3) and signal-to-noise ratio.}

\figcaption[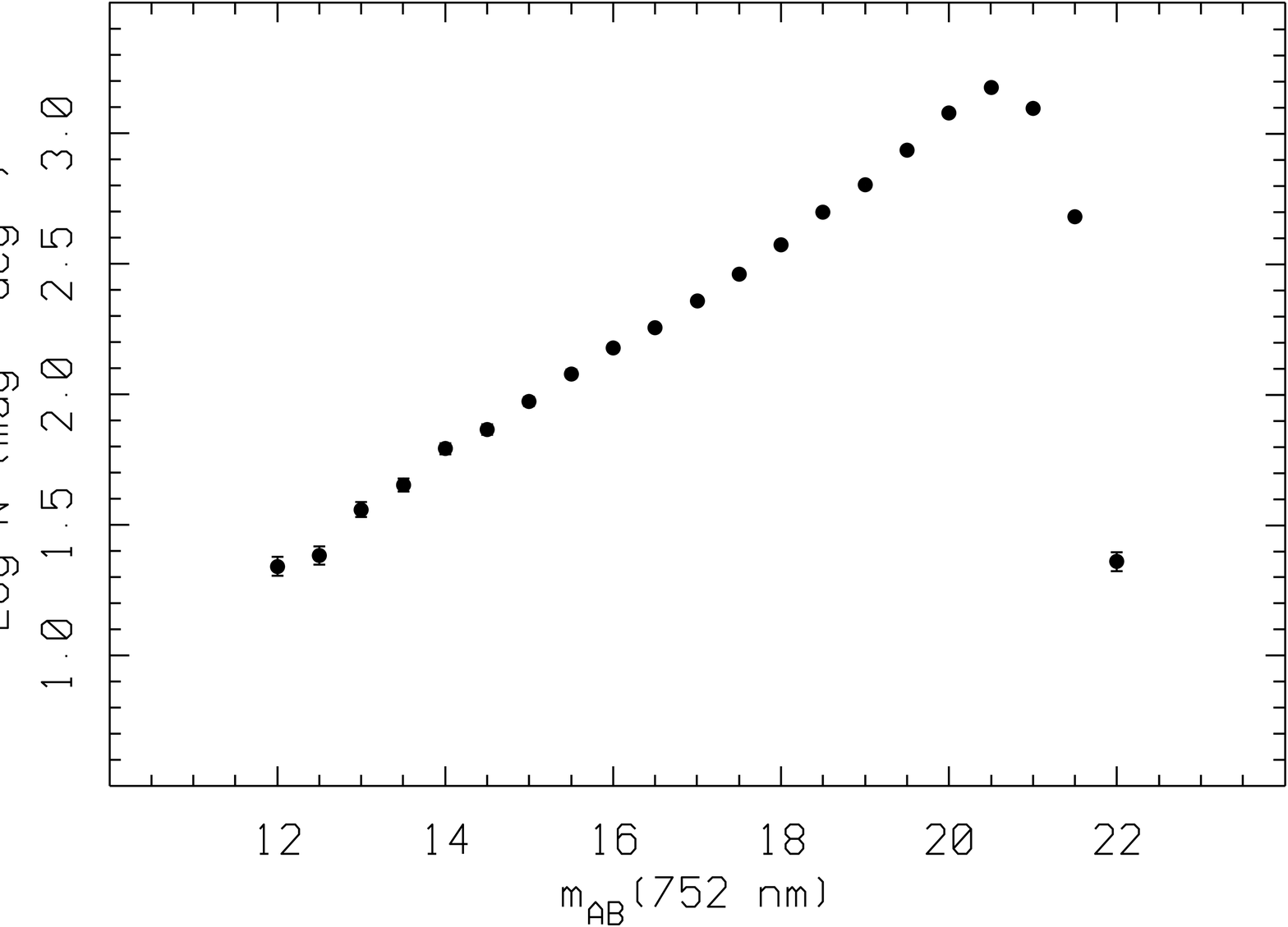]{Differential object counts. The number of detected 
objects, per square degree per magnitude interval, in the right ascension 
range 12-15 hrs is plotted vs. magnitude in the 752 nm band.}

\clearpage
\plotone{f1.eps}
\clearpage
\plotone{f2.eps}
\clearpage
\plotone{f3.eps}
\clearpage
\plotone{f4.eps}
\clearpage
\plotone{f5.eps}

\end{document}